\documentclass[allclo]{FBSart}
\usepackage{amsfonts}
\usepackage{amssymb}
\usepackage[dvips]{graphicx}

\newcommand{\subsize}{\scriptsize}
\newcommand{\sub}[1]{_\textrm{\subsize #1}}

\title{Condensates and correlated boson systems}

\author{O. S{\o}rensen\thanks{\textit{E-mail address:} 
 oles@phys.au.dk}, D. V. Fedorov, and A. S. Jensen}

\institute{Department of Physics and Astronomy, University of Aarhus,
  Denmark}

\runningauthor{O. S{\o}rensen, D. V. Fedorov, and A. S. Jensen}

\runningtitle{Correlated many-boson systems with arbitrary scattering
  length}

\begin{document}

\maketitle

\begin{abstract}
  We study two-body correlations in a many-boson system with a
  hyperspherical approach, where we can use arbitrary scattering
  length and include two-body bound states.  As a special application
  we look on Bose-Einstein condensation and calculate the stability
  criterium in a comparison with the experimental criterium and the
  theoretical criterium from the Gross-Pitaevskii equation.
\end{abstract}

\section{Introduction}

Bose-Einstein condensates (BEC) of alkali atoms, e.g., Rb and Na, have
succesfully been created in the last decade \cite{bec95}.  The JILA
experiments \cite{don01,don02} touch upon the collapse of boson
systems at large two-body scattering length $a_s$ due to the formation
of two-body bound states.  The stability criterium has recently been
measured to be $N|a_s|/b_0 < 0.55$ \cite{cla03}, where $N$ is the
number of bosons, $a_s$ is the two-body $s$-wave scattering length,
and $b_0^3 = \sqrt{\hbar/(m\bar\omega)}$ with the geometric mean
$\bar\omega$ of the angular frequencies of the deformed oscillator.
The properties of BEC have largely been accounted for by the
Gross-Pitaevskii equation (GPE) \cite{pet01,pit03} with a zero-range
interaction, which reproduces the correct average energy, density
distribution, and stability criterium \cite{gam01}.  This
approximation is usually succesful at low density $n$, where the
combination $n|a_s|^3$ is very small.

The need for a more realistic potential is becoming more important
with the cases of stronger attraction or larger densities, where
critical phenomena like collapse of a condensate \cite{don01} and the
conversion into molecular BEC occur \cite{don02}.  In addition, the
pure mean-field structure of the wave function is not able to account
for the correlations between the particles.  Attempts on top of the
mean-field have been tried, e.g., including pairing correlations
between the bosons \cite{kok02b} in the Hartree-Fock-Bogoliubov
formulation.  Alternatively the Jastrow ansatz with the many-body wave
function as a product of two-body amplitudes have lead to good results
for large scattering length \cite{cow01}.  Quantum Monte-Carlo
calculations have been applied for a large number of bosons in a
density matrix formulation \cite{kra96}, and more detailed studies for
smaller numbers \cite{cer00,blu01,ast03}.

Recently \cite{boh98} a hyperspherical description analogous to the
mean-field yielded the stability criterium $N|a_s|/b_0<0.67$ for a
spherical external field.  This was extended \cite{sor02col} to treat
two-body correlations with realistic finite-range potentials and
applied to a small number of particles.  Recently \cite{sor03col} this
was applied to a large number of particles and some universal scalings
were extracted from the results.  This paper will review these results
and discuss the various structure descriptions possible within the
model.

\section{Theory}

We study the $N$-boson system of identical, interacting bosons of mass
$m$ trapped by an isotropic harmonic external field of angular
frequency $\omega$.  The total Hamiltonian is then
\begin{eqnarray}
  \hat H\sub{total}=
  \sum_{i=1}^N \frac{\hat p_i^2}{2m}
  +\sum_{i=1}^N\frac12m\omega^2r_i^2
  +\sum_{i<j=1}^NV(r_{ij})
  \;.
\end{eqnarray}
We focus on the relative degrees of freedom and use hyperspherical
coordinates \cite{sor02col}: hyperradius $\rho$ given by $\rho^2 =
\sum_{i<j}r_{ij}^2/N$, where $r_{ij} = |\vec r_i-\vec r_j|$, and
hyperangles $\Omega$ relating to the remaining $3N-4$ relative degrees
of freedom.  Due to the properties of the harmonic oscillator, the
center-of-mass motion separates out, leaving a relative eigenvalue
equation:
\begin{eqnarray}
  (\hat H\sub{total}-\hat H\sub{c.m.}-E)\Psi(\rho,\Omega)=0
  \;.
\end{eqnarray}
We factorize the relative wave function as
$\Psi(\rho,\Omega)=\rho^{-(3N-4)/2}f(\rho)\Phi(\rho,\Omega)$.  This
way we obtain an effective radial equation for $f(\rho)$ as the
eigenfunction of energy $E$ in the effective potential $U(\rho)$
\begin{eqnarray}
  &&
  \bigg(-\frac{\hbar^2}{2m}\frac{d^2}{d\rho^2}+U(\rho)-E\bigg)f(\rho)=0
  \;,\\
  &&
  \frac{2mU(\rho)}{\hbar^2}
  =
  \frac{(3N-4)(3N-6)}{4\rho^2}
  +\frac{\lambda(\rho)}{\rho^2}
  +\frac{\rho^2}{b\sub t^4}
  \;.
\end{eqnarray}
Here $b\sub t=\sqrt{\hbar/(m\omega)}$ is the length unit of a harmonic
trapping potential of frequency $\nu=\omega/(2\pi)$.  The
$\rho$-dependent $\lambda$ is an effective angular potential, which
includes the effects of interactions and correlations.

Assuming then a wave function which is basically a sum, and not a
product, of two-body amplitudes as
\begin{eqnarray}
  \Phi(\rho,\Omega) 
  =
  \sum_{i<j}\phi(\rho,r_{ij})
\end{eqnarray}
we are able to solve the angular part of the many-body Schr\"odinger
equation as just a one-dimensional integro-differential equation with
up to two-dimensional integrals.  In the angular equation we use a
finite-range Gaussian $V(r_{ij}) = V_0\exp(-r_{ij}^2/b^2)$ as the
two-body interaction potential.  The only additional approximation
consists in assuming that the range $b$ of this potential is much
smaller than the average distance between the particles.  The
scattering length $a_s$ can assume any value, and there can be any
number of two-body bound states, at least in principle; in actual
calculations the accuracy is better with as few two-body bound states
as possible.  The usual zero-range interaction used in the mean-field
amounts to equating $a_s$ with the Born-approximation $a\sub B$ to the
scattering length.  In comparison we generally have that $a_s$ and
$a\sub B$ differ substantially.  The Born-approximation is given by
\begin{eqnarray}
  a\sub B
  \equiv
  \frac{m}{\hbar^2}\int_0^\infty dr\; r^2 V(r)
  \;.
\end{eqnarray}

When we assume that $\Phi(\rho,\Omega)$ is independent of angular
coordinates, the angular potential becomes for $N\gg1$
\begin{eqnarray}
  \lambda_\delta(\rho)
  =
  \frac32\sqrt{\frac3\pi}N^{7/2}\frac{a\sub B}\rho
  \;\to\;
  \frac32\sqrt{\frac3\pi}N^{7/2}\frac{a_s}\rho
  \;,
\end{eqnarray}
where we due to the mean-field like assumption of no internal
structure replaced $a\sub B$ by $a_s$.  This is generally the
asymptotic angular eigenvalue for the lowest state above two-body
bound states.  The mean-field description only contains this effective
interaction potential.  Two-body structures and bound states in this
two-body correlated model is clearly beyond the mean-field.

\section{Results}

\begin{figure}[htb]
  \centering 
  \input{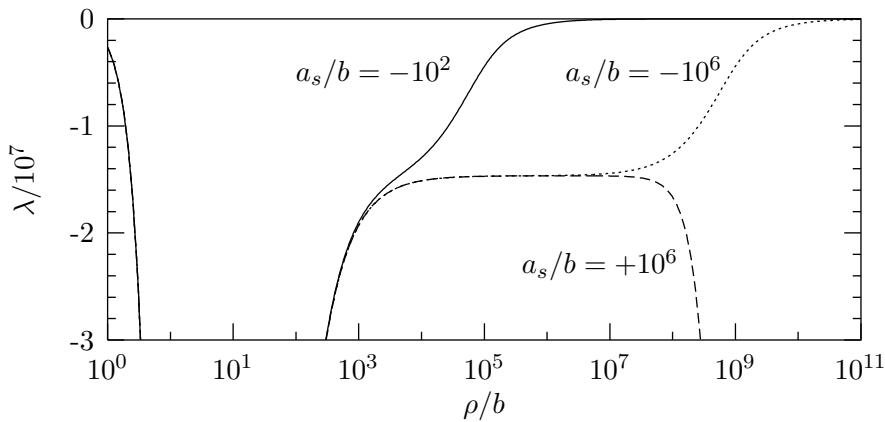}
  \caption[]
  {The angular eigenvalue  $\lambda$ for $N=1000$ and a Gaussian
    interaction potential of range $b$.  The hyperradius $\rho$ is in
    units of $b$ and the two-body $s$-wave scattering lengths $a_s$ is
    indicated as $a_s/b$ on the plot.}
  \label{fig:trento1}
\end{figure}

With a Gaussian two-body interaction potential of range $b$ the
qualitative features of $\lambda$ are as shown in
figure~\ref{fig:trento1}.  The large-$\rho$ features only depend on
$a_s$ and therefore $b$ only matters for the model-dependent region at
small $\rho$ ($\rho/b < 10^3$ in figure~\ref{fig:trento1}).  At small
and negative scattering length the angular eigenvalue quickly
approaches zero as $1.5N^{7/2}a_s/\rho$.  At larger, but still
negative, scattering length the angular eigenvalue follows a constant
value $-1.6N^{7/3}$ before approaching zero as before.  In the case of
a bound two-body state, when the scattering length turns positive,
$\lambda$ diverges as $-2\rho^2/a_s^2$.  Thus a description of
two-body bound states is also possible within the model.

\begin{figure}[htb]
  \centering
  \input{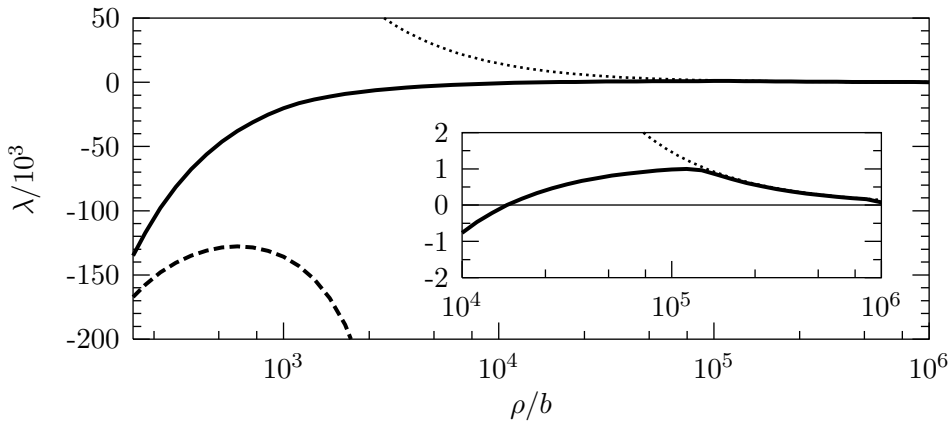}
  \caption[]
  {The two lowest angular eigenvalues (dashed and solid curves) for
    $N=100$, $a_s/b=+10$, and one bound two-body state.  The dotted
    curve is $\lambda_\delta$ for the same scattering length.}
  \label{fig:trento2}
\end{figure}

The inclusion of the two-body bound state is evident from considering
the full spectrum of angular eigenvalues for the case of, e.g., one
two-body bound state.  In figure~\ref{fig:trento2} we show the two
lowest angular eigenvalues in such a case with one two-body bound
state and positive scattering length.  The lowest eigenvalue (dashed
curve) diverges to minus infinity proportional to $\rho^2$.  This
corresponds to the bound state.  The second eigenvalue (solid curve)
is negative at small hyperradii but turns positive at larger and
approaches the asymptotic behavior $\lambda\propto a_s/\rho$ (dotted
curve, see inset).

\begin{figure}[htb]
  \centering
  \input{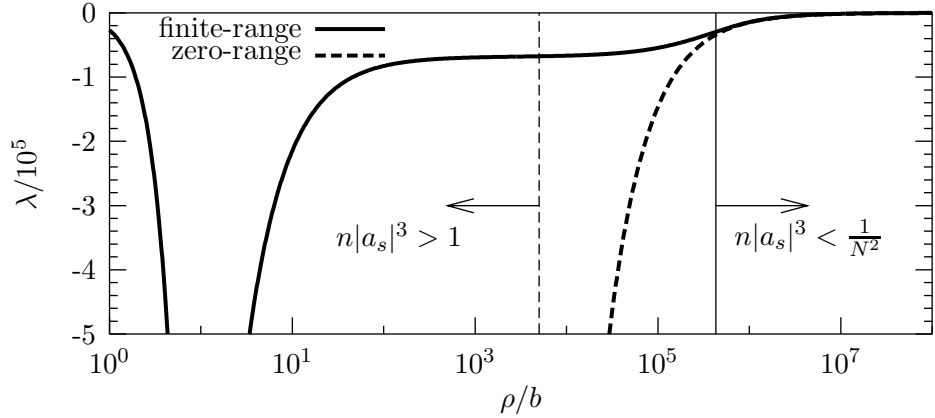}
  \caption[]
  {The lowest eigenvalue for $a_s/b=-10^4$, $N=100$, and no bound
    two-body states (solid).  The dashed curve is $\lambda_\delta$ for
    the same scattering length.  The vertical lines indicate regions
    of different density.}
  \label{fig:trento3}
\end{figure}

The deviations of this method from the mean-field are illustrated in
figure~\ref{fig:trento3}, where the lowest angular eigenvalue for a
case with no two-body bound state and negative scattering length is
compared to the zero-range angular eigenvalue for the same scattering
length.  For low density $n|a_s|^3<1/N^2$ the effective energy of the
two methods coincide, for larger densities the mean-field energy
diverges, while the energy from the finite-range model remains finite.
Moreover, it deviates in a region where the density is still
relatively low $n|a_s|^3<1$ so higher-order correlations (especially
three-body) do not play a role yet.

\begin{figure}[htb]
  \centering
  \input{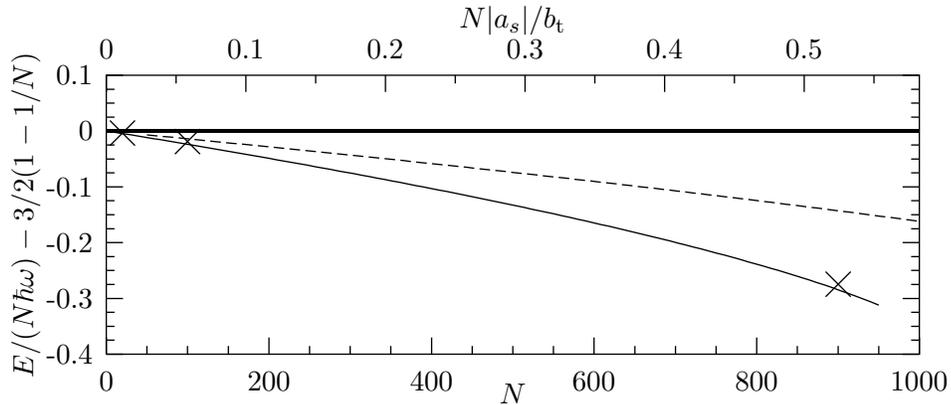}
  \caption[]
  {Interaction energy per particle for $b\sub t/b=1442$, $a_s/b=-0.84$
    ($a\sub B/b=-0.5$) calculated by the GPE (solid line) and the
    two-body correlated method (crosses).  The dashed line is the GPE
    calculation with $a_s=a\sub B=-0.5b$.}
  \label{fig:trento4}
\end{figure}

A direct comparison of the interaction energy is done by calculating
the interaction energy per particle for the zero-range mean-field and
for this finite-range model with correlations, see
figure~\ref{fig:trento4}.  The solid curve shows the GPE interaction
energy, which increases in magnitude until there is no stable system
for $N|a_s|/b_0>0.55$.  The crosses are the results from this
finite-range correlated model, which almost coincide with the GPE
results.  Setting the barrier height of the radial potential with
$\lambda=\lambda_\delta$ equal to the oscillator energy yields the
stability criterium $N|a_s|/b_0<0.53$, which agrees well with the
above.  The dashed curve is the result for the mean-field with a
scattering length equal to the Born-approximation of the finite-range
Gaussian we used in the calculation.  The deviation of the two GPE
calculations, is equivalent to the observation that the mean-field
model with a zero-range interaction gives the correct result, but not
with a finite-range interaction.  However, the correlated model
reproduces the correct interaction energy with a finite-range
interaction of the true scattering length.  We interpret this as a
confirmation that the crucial degrees of freedom are included in the
ansatz for the wave function.

\section{Conclusion}

We presented the key results of a hyperspherical study of two-body
correlations in BEC.  We confirmed the stability criterium.  Moreover,
structure beyond the mean-field is observed due to the description of
two-body bound states and the effective energy at larger densities.
We expect this model in the future allows more detailed analysis of
the coupling between molecular BEC and atomic BEC.  A possible
extension is the inclusion of the important three-body correlations
for a study of the recombination process in dilute atomic medium.


\end{document}